%% file: paper_v3.tex
\documentclass[aps,prl,twocolumn,superscriptaddress,numerical,]{revtex4-2}
\usepackage[utf8]{inputenc}
\usepackage[english]{babel}
\usepackage[T1]{fontenc}
\usepackage{amsmath,amssymb,dsfont,lipsum}
\usepackage{tikz}
\usepackage{lipsum}
\usepackage{dsfont}
\usepackage{physics}
\usepackage{float}
\usepackage[colorlinks,
citecolor=purple,
linkcolor=purple,
urlcolor=purple]{hyperref}
\usepackage{braket}
\usepackage{orcidlink}
\usepackage{multirow}
\graphicspath{{images}}

\renewcommand{\section}[1]{\textbf{\emph{#1}}.\!}
\renewcommand{\subsection}[1]{{\emph{#1}}.---}

\begin{document}

\title{Anomalous Diffusion and Superdiffusion in Integrable Spin Chains \\ via a Soliton Gas Mapping}

\author{Andrew Urilyon~\orcidlink{0000-0001-8960-5388}}
\affiliation{Laboratoire de Physique Th\'eorique et Mod\'elisation, CNRS UMR 8089, CY Cergy Paris Universit\'e, 95302 Cergy-Pontoise Cedex, France}
\affiliation{JEIP, UAR 3573 CNRS, Collège de France, PSL Research University,
11 Place Marcelin Berthelot, 75321 Paris Cedex 05, France}

\author{Romain Vasseur~\orcidlink{0000-0002-4636-4139}}
\affiliation{Department of Theoretical Physics, University of Geneva, 24 quai Ernest-Ansermet, 1211 Geneva, Switzerland}

 \author{Sarang Gopalakrishnan~\orcidlink{0000-0002-7493-7600}}
\affiliation{Department of Electrical and Computer Engineering, Princeton University, Princeton, NJ 08544, USA}

\author{Jacopo De Nardis~\orcidlink{0000-0001-7877-0329}}
\affiliation{Laboratoire de Physique Th\'eorique et Mod\'elisation, CNRS UMR 8089, CY Cergy Paris Universit\'e, 95302 Cergy-Pontoise Cedex, France}
\affiliation{JEIP, UAR 3573 CNRS, Collège de France, PSL Research University,
11 Place Marcelin Berthelot, 75321 Paris Cedex 05, France}

\begin{abstract}
We introduce a multi-species soliton gas in which each species has a distinct effective scattering length, and the repulsive scattering shift is set by the smaller of the two colliding particles.
We argue that this model shares key quasiparticle and scattering features with the XXZ spin chain.
We show that fixing only the functional decay of bare velocities with particle size is sufficient to reproduce the XXZ spin-transport phase diagram: diffusion (with anomalous fluctuations) in the anisotropic regime and superdiffusion at the isotropic point.
We then demonstrate that the statistics of \textit{charge transfer} differs qualitatively from that of \textit{particle trajectories}.
For large particles, trajectories are Gaussian in the diffusive regime and appear to exhibit KPZ statistics at the isotropic point, providing a direct microscopic signature of KPZ physics in integrable quasiparticle motion.
In contrast, charge-transfer fluctuations are anomalous in the anisotropic regime, while they cross over to Gaussian statistics at late times at the isotropic point, reconciling non-Gaussian
trajectory fluctuations with Gaussian charge-transfer statistics.
Our results establish this soliton gas as a minimal framework for anomalous spin and charge transport in integrable systems, and offer new insight into the origin of KPZ fluctuations in isotropic integrable models.
\end{abstract}
\maketitle

\section{Introduction} 
Strongly interacting many-body systems are notoriously difficult to describe, especially out of equilibrium, yet they can display behavior qualitatively distinct from weakly interacting or dilute limits. Transport theory provides a central arena where such differences emerge. While transport is often \emph{normal}, i.e. diffusive, robust exceptions arise in low-dimensional systems.

A prominent example is the anomalous transport of spin or charge in one-dimensional quantum systems \cite{PhysRevX.6.041065, PhysRevLett.117.207201,PhysRevLett.122.090601,PhysRevLett.119.195301, PhysRevLett.121.160603, de_nardis_diffusion_2019, doyon_lecture_2019, doyon2021free, SciPostPhys.3.6.039, doyon2022diffusion, PhysRevLett.127.130601, Doyon2018, PhysRevB.90.161101,Bastianello_2022, 10.21468/SciPostPhysCore.3.2.016, bastianello2018sinh, RevModPhys.93.025003, PhysRevLett.122.240606, Bulchandani_2021, Hubner2023, PhysRevResearch.6.013328,doi:10.1126/science.abf0147,10.21468/SciPostPhysCore.7.2.025,PhysRevLett.124.140603,Besse2023DissipativeGHD, Bulchandani2018BetheBoltzmann, Moller2022Bridging,PhysRevLett.123.130602, Moller2024Anomalous,10.21468/SciPostPhys.9.4.044,Doyon2025Perspective,Bonnemain2022,PhysRevLett.133.107102,PhysRevLett.133.113402,PhysRevLett.126.090602,Kao2021,Le2023,PhysRevX.12.041032,2406.17569,Wei2022,2505.10550,dubois2025}. Integrable systems are usually associated with ballistic transport \cite{ZotosNaefPrelovsek1997,Zotos1999Drude,Prosen2011NESS,Prosen2013Quasilocal,IlievskiDeNardis2017HubbardHydro,UrichukOezKlumperSirker2019SciPostDrude,KlumperSakai2019FiniteSizeDrude,Karrasch2013DrudeWeight,HeidrichMeisner2005ThermalDrude}, whereas non-integrable one-dimensional systems with momentum conservation often exhibit anomalous hydrodynamic tails in Kardar--Parisi--Zhang (KPZ) or L\'evy universality classes \cite{Spohn2014NLFHAnharmonic,PhysRevA.92.043612,PhysRevLett.124.236802,PhysRevB.73.035113}. It was therefore surprising that \emph{isotropic} integrable spin chains can display KPZ fluctuations \cite{Bulchandani_review,LjubotinaZnidaricProsen2019KPZ,GopalakrishnanVasseur2019KT,PhysRevB.101.041411,DeNardisGopalakrishnanIlievskiVasseur2020Solitons,
IlievskiDeNardisGopalakrishnanVasseurWare2021Superuniversality,DeNardisGopalakrishnanVasseur2023NLFH,
WeiRubioAbadalYeMachadoKempSrakaewHollerithRuiGopalakrishnanYaoBlochZeiher2022Science,YeMachadoKempHutsonYao2022Universal,PhysRevLett.134.097104,Prosen2026,krajnik2020_kpz,
PhysRevE.106.L062202,PhysRevB.105.L100403,PhysRevLett.133.256301,PhysRevB.110.L180301,2506.12133,2510.26897}, a phenomenon now also confirmed experimentally \cite{WeiRubioAbadal2022Science,ScheieShermanDupontNaglerStoneGranrothMooreTennant2021NatPhys,RosenbergAndersen2024Science}.

\begin{figure}[!b]
  \centering
  \includegraphics[width=1\linewidth]{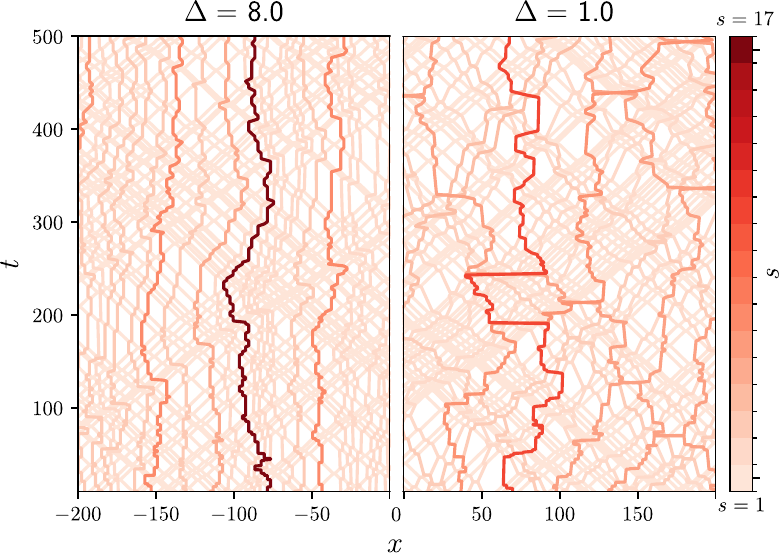}  
  \caption{Particle trajectories for $\Delta = 8$ and $\Delta = 1$ for different sizes $s$. Because of the different bare-velocity scaling, large particles at $\Delta = 8$ move mainly through collisions with light ones, leading to random-walk-like behavior, whereas at $\Delta = 1$ they undergo rare jumps of order $O(s)$ due to collisions with other large particles, producing long-time tails. Note that the largest visible jumps are due to the cumulative jump due to multiple particle scattering events.}
  \label{fig:example_trajectory}
\end{figure}

Despite intense activity, a complete microscopic explanation for KPZ scaling in integrable spin chains is still lacking: current understanding relies mainly on scaling arguments fixing the dynamical exponent \cite{GopalakrishnanVasseur2019KT} and on partial nonlinear fluctuating hydrodynamic descriptions \cite{DeNardisGopalakrishnanVasseur2023NLFH,Takeuchi2025}. Moreover, away from isotropy integrable models can exhibit another type of anomalous transport, namely diffusion with anomalous fluctuations~\cite{DeNardisBernardDoyon2018PRL,DeNardisBernardDoyon2019SciPost,GopalakrishnanHuseKhemaniVasseur2018PRB,Krajnik2022,PhysRevLett.128.090604,GopalakrishnanMcCullochVasseur2024PNAS,mcculloch2024,krajnik2024_dynCrit,yoshimura_anomalous_2024}. In the XXZ chain this is understood in terms of the interplay between ballistically propagating magnons and their bound states: the effective spin carried by fast quasiparticles is dynamically screened through interactions with large bound states \cite{Gopalakrishnan0-v2023}, while scattering-induced shifts generate anomalous diffusive fluctuations. How these mechanisms reorganize at the isotropic point into KPZ-like behavior remains mysterious.

\section{Summary of results}
We introduce and analyze a minimal \emph{classical} proxy for XXZ quasiparticle dynamics: a one-dimensional multi-species gas~\cite{Jepsen1965HardRods,PhysRev.171.224,Percus_HR_1969,Boldrighini1983,Spohn1991,Boldrighini1997, PhysRevE.104.064124,PhysRevE.108.064130,2504.09201,b587-8yyt,Ferrari2025} in which species represent bound states ("strings") of different lengths, each with two chiralities and interacting through a prescribed cumulative scattering shift; see Eqs.~\eqref{eq:scatteringT} and~\eqref{eq:traje}. Despite its simplicity, the model reproduces several key features of spin transport in the Heisenberg chain.

\begin{figure}[t!]
  \centering
    \includegraphics[width=1\linewidth]{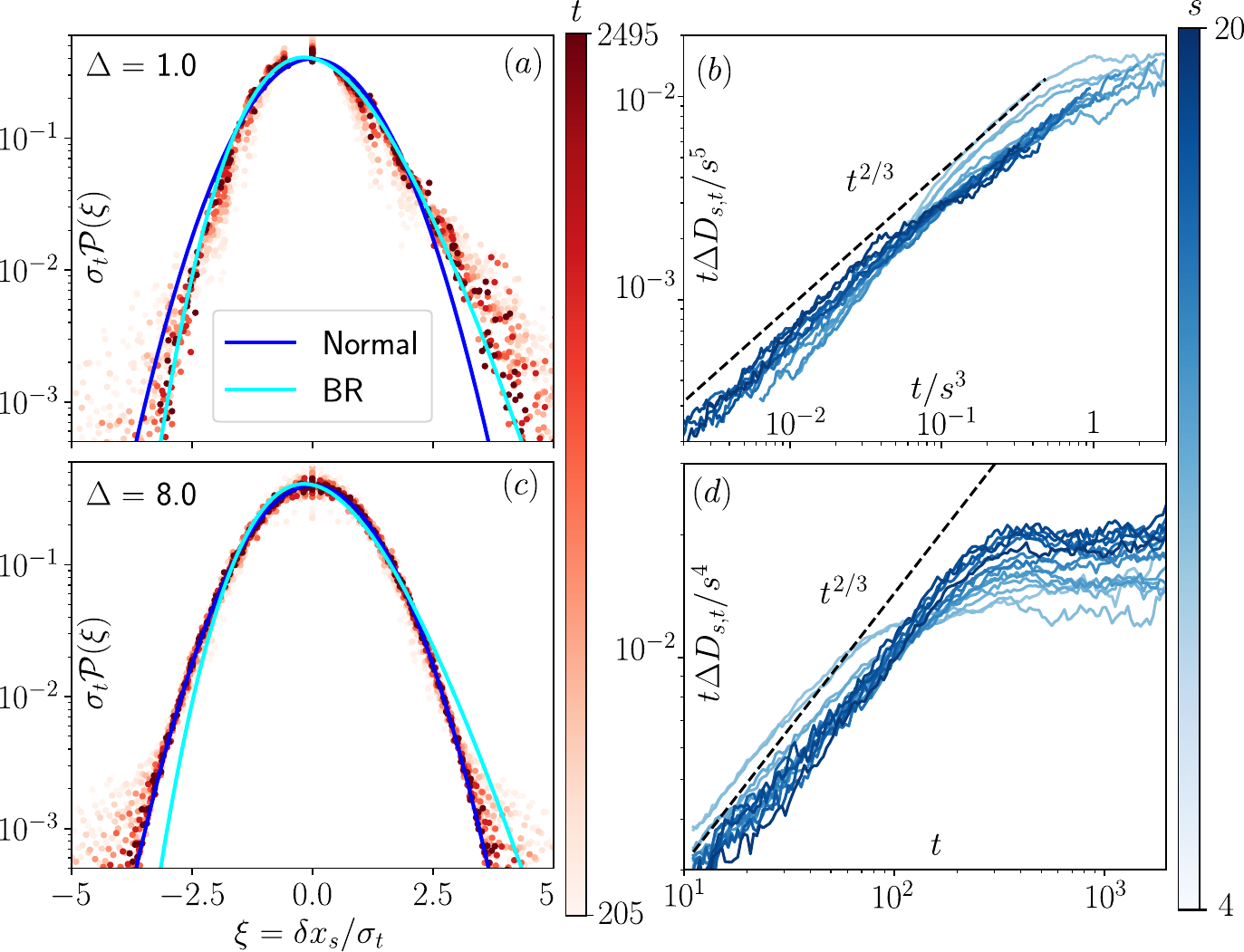}
  \caption{Distribution of trajectory fluctuations $\mathcal{P}(\delta x_{s})$ in Eq.~\eqref{eq:displacement} for right-moving particles $\sigma=1$ (also averaged with left-movers under $\xi \to -\xi$), compared with a Gaussian and a skewed KPZ (Baik--Rains) distribution (see also Fig. \ref{fig:smallStringPDF} for smaller $s$ values). All curves are rescaled to zero mean and unit variance (the latter indicated by $\sigma_t$). The right panel shows the approach to asymptotic diffusion $t\Delta D_{s,t}=t(D_s(t)-D_s)$, plotted against the rescaled time $t/s^3$ for $\Delta=1$ and against $t$ for $\Delta=8$. }
  \label{fig:asymptotic_trajectory}
\end{figure}

\begin{figure}[t!]
  \centering
    \includegraphics[width=1\linewidth]{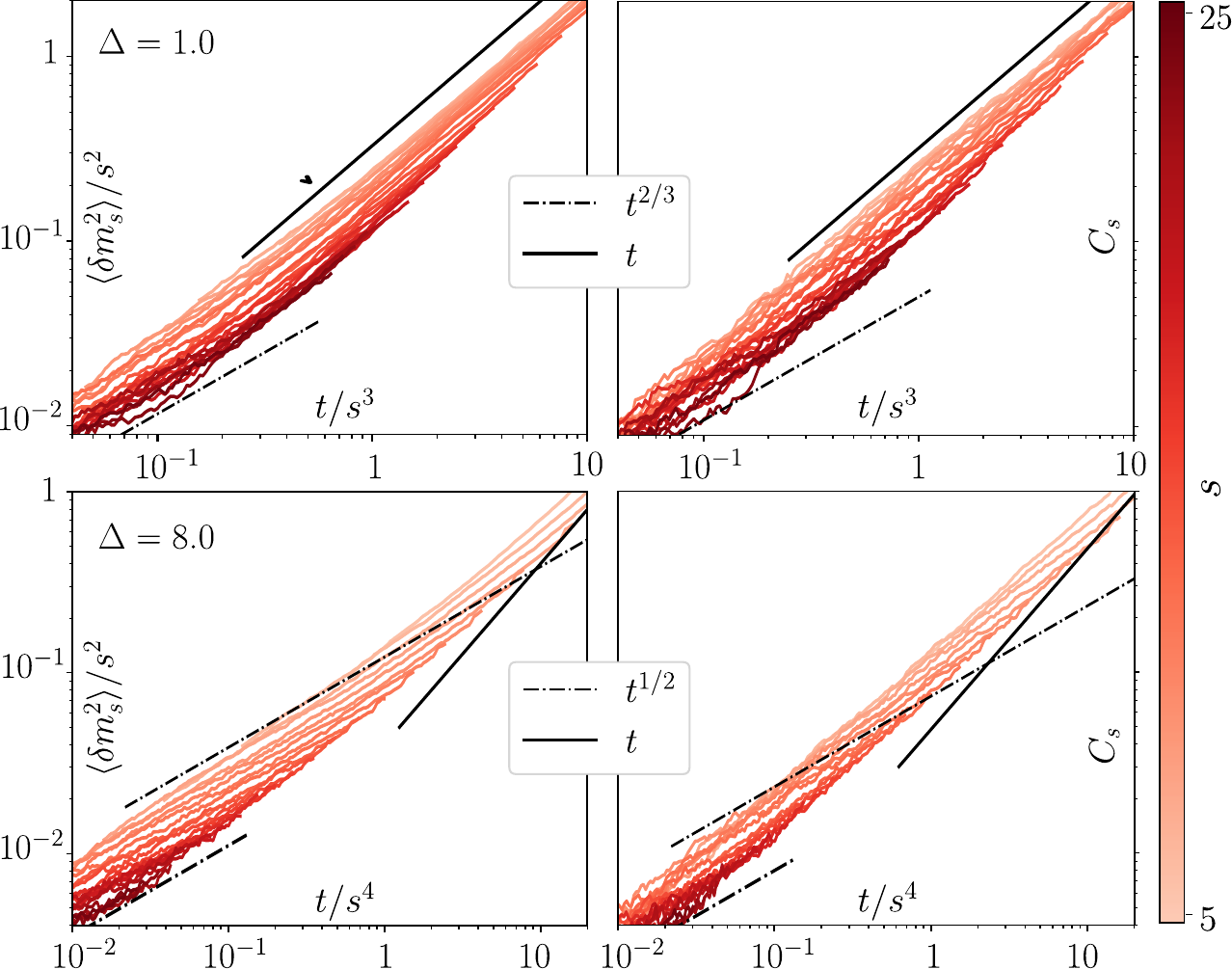}\\
  \caption{Behavior of the magnetization transfer for each species (see Eq.~\eqref{eq:dm-def}), $\langle \delta m _s^2 \rangle$ (left) and the sign-flipped off-diagonal sum $C_s = -\sum_{s' \neq s} \langle \delta m_s \delta m_{s'} \rangle / s^2$ (right) for both $\Delta = 1$ and $\Delta = 8$. Different red curves correspond to different species $s$. The data demonstrate the presence of a transition from superdiffusive to ballistic behavior for $\Delta=1$ (top) and from diffusive to ballistic for $\Delta>1$ (bottom).}
  \label{fig:screening_magnetization}
\end{figure}

At the isotropic point, large particles undergo rare but large displacements due to collisions with other large particles (Fig.~\ref{fig:example_trajectory}), producing a crossover in \emph{trajectory} fluctuations from KPZ-like to diffusive behavior (Fig.~\ref{fig:asymptotic_trajectory}) and leading to superdiffusive \emph{charge} transport with asymptotically Gaussian charge fluctuations (Fig.~\ref{fig:magnetization_distribution}), consistent with numerics and experiments~\cite{RosenbergAndersen2024Science,PhysRevLett.128.090604,PhysRevB.108.235139,f3c4-n21z,krajnik2024_dynCrit}. Away from isotropy, by contrast, large particles are effectively immobilized and move mainly through collisions with ballistically propagating light particles, leading to a single-file diffusion picture. Importantly, all quasiparticles carry positive charge: screening and anomalous diffusion arise not from cancellations between opposite carriers, but from strong dynamical inter-species correlations (Eq.~\eqref{eq:dm2-decomp} and Fig.~\ref{fig:screening_magnetization}). Overall, the model provides a microscopic classical description of anomalous diffusion and superdiffusion in the Heisenberg chain, while giving direct access to \emph{single-quasiparticle trajectories}.

\section{The multi-species soliton gas}
\label{sec:ms-hrg}
The spin-$\tfrac12$ XXZ chain is defined by
\begin{equation}
    H=\sum_{j}\Big(S^x_j S^x_{j+1}+S^y_j S^y_{j+1}+\Delta\, S^z_j S^z_{j+1}\Big)\,.
\end{equation}
Its thermodynamic spectrum contains magnons and bound states (strings) labeled by size $s\ge1$ and rapidity $\theta$ \cite{Bethe1931,Orbach1958,Takahashi1971,Takahashi1999Book}. For $|\Delta|\ge1$ there is no upper bound on $s$, and thermal states populate arbitrarily large strings. The main distinction between the isotropic point $\Delta=1$ and the anisotropic regime $\Delta>1$ is the large-$s$ scaling of effective velocities: algebraic at $\Delta=1$ and exponential at $\Delta>1$. To isolate the minimal ingredient behind this transport phenomenology, we introduce a classical integrable proxy: a multi-species gas whose key input is precisely this $s$-dependence of velocities.

We consider species $s=1,\dots,s_{\max}$ with chiralities $\sigma=\pm1$, bare velocities
\[
v^\sigma_s=\sigma \frac{e^{-(s-1)\eta}}{s^2}, \qquad \eta={\rm arccosh}(\Delta),
\]
and equilibrium densities $\overline\rho_s=1/[s(s+1)(s+2)]$, chosen so that $\sum_{s\ge1} s\bar\rho_s=1/2$. Thus $\eta\ge0$ interpolates between algebraic decay ($\eta=0$) and exponential suppression ($\eta>0$).

We prescribe the nonlinear scattering length
\begin{equation}
\mathfrak{a}_{s,s'}=2\min(s,s')-\delta_{s,s'}\,,
    \label{eq:scatteringT}
\end{equation}
which mimics the fusion hierarchy and scattering shift in the strongly anisotropic XXZ chain \cite{Klumper1993,IlievskiQuinnDeNardisBrockmann2016}. The shift is controlled by the smaller of the two particle sizes, as in soliton gases such as box-ball system \cite{TakahashiSatsuma1990,KunibaMisguichPasquier2020,CroydonKatoSasadaTsujimoto2019,FerrariNguyenRollaWang2019}, and is therefore \textit{not linear} in the sizes $s,s'$. For a single species Eq.~\eqref{eq:scatteringT} reduces to the usual hard-rod shift, but for multiple species it is not the geometric multi-species hard-rod gas. Although Eq.~\eqref{eq:scatteringT} is not the exact XXZ phase shift at generic $\Delta$, it captures the large-string scattering structure that we identify as responsible for the universal dynamical consequences of many interacting species with a broad velocity distribution.

The physical intuition behind Eq.~\eqref{eq:scatteringT} is tied to magnetization screening \cite{ganahl2013quantumbowlingparticleholetransmutation,GopalakrishnanVasseur2019KT,Gopalakrishnan0-v2023,DeNardis2022,PhysRevLett.127.057201}. In the large-$\Delta$ XXZ chain, small bound states crossing large ones repeatedly flip their internal orientation, leading to an effective randomization of the charge they carry. In our classical proxy, the same screening mechanism is implemented kinematically through collision-induced displacements.

All particles carry a positive charge proportional to their size, so the local magnetization density is
\begin{equation}
m(x,t)=\lim_{s_{\max} \to \infty}\sum_{s=1}^{s_{\max}} s\,{\rho}_s(x,t)\,-1/2 ,
    \label{eq:magnetization}
\end{equation}
with ${\rho}_s(x,t)$ the density of species $s$ within an interval of size $dx$ at position $x$. Interactions dress the ballistic motion in a nontrivial way. The microscopic dynamics is defined by free motion in auxiliary coordinates together with cumulative two-body scattering shifts. Wherever free trajectories cross the interacting particle positions are scattered. The trajectory $x_{s,i}(t)$ of the $i$-th particle of species $s$ and chirality $\sigma_i$ satisfies
\begin{equation}
    v^{\sigma_i}_{s}\, t + X^{(0)}_{s,i}
    =
    x_{s,i}(t)
    +\frac12\sum_{s',j}\mathfrak{a}_{s,s'}\,
    \mathrm{sign}\!\big(x_{s,i}(t)-x_{s',j}(t)\big)\,,
    \label{eq:traje}
\end{equation}
where $X^{(0)}_{s,i}$ are stochastic auxiliary initial positions sampled as described in the End Matter. Equation~\eqref{eq:traje} should be read as a ledger of ordering changes: a shift is accumulated when the corresponding free trajectories cross, so the scattering events remain local even though the implicit formula is written in terms of the full ordering. Equations of this form can be inverted through a convex minimization principle and provide an exact microscopic definition of the interacting soliton gas \cite{10.21468/SciPostPhys.13.3.072,Bonnemain2025,2503.08018,Hubner2023,PhysRevLett.132.251602}.

\section{Particle trajectories}
\label{sec:trajsec}
The dynamics generated by Eq.~\eqref{eq:traje} is analogous to other integrable soliton gases with factorized elastic scattering: particles retain their bare velocities in the auxiliary coordinates, while interactions generate cumulative position shifts. Coarse graining over many collisions, we characterize a tagged particle through its displacement
\begin{equation}\label{eq:displacement}
    \delta x_{s,i}(t)=x_{s,i}(t)-x_{s,i}(0)\,.
\end{equation}
Its effective velocity is the long-time mean drift,
\begin{equation}
    \bigl(v^{\sigma}_{s}\bigr)^{\rm eff}
    =\lim_{t\to\infty}\frac{1}{t}\,\Big\langle \delta x_{s,i}(t)\Big\rangle
    =\sigma \frac{ (s+2) e^{-\eta (s-1)}}{s (s+1)} ,
    \label{eq:veff-def}
\end{equation}
which scales as $s^{-1}$ at $\Delta=1$, as in the XXX model.

Fluctuations around the mean drift define an effective \textit{diffusion constant for particle trajectories}, which can be evaluated at a given time $t$,
\begin{equation}\label{eq:finitetimediff}
      D_s(t) \equiv  \frac{d}{dt} \Big[  \,
    \Big\langle\bigl(\delta x_{s,i}(t)-\langle \delta x_{s,i}(t)\rangle\bigr)^2\Big\rangle\Big],
\end{equation}
and whose infinite-time limit $D_s = \lim_{t \to \infty} D_s(t)$ is \textit{finite} for each fixed species $s$: in an interacting integrable gas, a tagged particle experiences an effectively random sequence of collisions at long times, which produces diffusive broadening of its trajectory.
At first sight this seems paradoxical: \textit{how can individual quasiparticle trajectories be diffusive, while the transported conserved charge displays superdiffusive spreading?}
The resolution is that \textit{large particles can  build up anomalously large fluctuations before crossing over to normal diffusion at parametrically large time-scales}. The latter is directly related to the scaling of the diffusion constant with $s$.  Indeed, in the anisotropic regime $\Delta>1$, the diffusion constants $D_s$ rapidly saturate to an $s$-independent value $D_\infty$, while at the isotropic point, one finds linear growth with $s$.
An explicit evaluation yields (see End Matter for more details),
\begin{eqnarray}
    D_s
    &=&
    \frac{8}{9}\sum_{s'=1}^{\infty}
    \min\!\left(
        1,\,
        \frac{(s+2)^3 s(s+1)}{(s'+2)^3 s'(s'+1)}
    \right)
    \left(\frac{s'+1}{s'}\right)^{2}
    \nonumber\\
    &&\hspace{1.8cm}
    -\frac{2(1+4s)}{3s^{2}}
    \;\xrightarrow[s\to\infty]{}\;
    \frac{10}{9}\,s\,.
    \label{eq:Ds-asympt}
\end{eqnarray}

To then characterize particle trajectories, we introduce the scaling functions $H_{ \Delta }(z)$ with asymptotics given by $H_{ \Delta }(z \ll 1) \sim z^{-1/6}$ and $H_{ \Delta }(z \gg 1) = {\rm const}$.
\textit{For $\Delta>1$} (see Fig.~\ref{fig:asymptotic_trajectory}(d)), the tagged-particle position admits the following form:
\begin{equation}
    x_{s,i}(t)
    =
    \bigl(v^\sigma_s\bigr)^{\rm eff} t
    +\sqrt{D_s}\ t^{1/2}\, H_{\rm \Delta >1}(t) \xi_{\rm G}
    +O(t^{0} s^2),
    \label{eq:traj-diff-aniso}
\end{equation}
with $\xi_{\rm G}$ a mean-zero, unit-variance normal (up to finite-time effects) random variable and $D_{s \gg 1} = D_\infty$.
\textit{At the isotropic point $\Delta=1$}, by contrast, the crossover time scales as $t_{*}(s)\sim s^{3}$, such that, see Fig.~\ref{fig:asymptotic_trajectory}(b):
\begin{equation}\label{eq:traj-diff-iso}
    x_{s,i}(t) \simeq  (v^\sigma_s\bigr)^{\rm eff} t + \sqrt{ \kappa_{\rm G} s t} \ H_{\Delta =1}(t/s^3) \ \xi^\sigma_{\rm G - KPZ} + O(t^0 s ) ,   
\end{equation}
where the unit-variance random variable $\xi^\sigma_{\rm G - KPZ}$ crosses over from a \textit{skewed} (whose sign is given by $\sigma$) random variable for $t/s^3 \ll 1$ (see Fig.~\ref{fig:asymptotic_trajectory}(a)), compatible with the Baik--Rains (BR) distribution characterizing stationary growing-interface fluctuations in the KPZ equation~\cite{KardarParisiZhang1986,TracyWidom1994Airy,PraehoferSpohn2002,Corwin2012}, to a normal Gaussian one for $t/s^3 \gg 1$.
Namely, trajectory fluctuations around their mean exhibit a crossover from KPZ to normal dynamics, similarly (yet opposite in time) to crossovers between different statistics known for the KPZ equation and the WASEP model~\cite{Sasamoto2010,LeDoussal2017}.

The origin of such a crossover is manifested at the level of a single trajectory (see Fig.~\ref{fig:example_trajectory}).
For $\Delta>1$, the bare velocities of large-$s$ particles are exponentially suppressed, and heavy particles are effectively immobile; their motion is dominated by frequent collisions with light species with random left/right chiralities, resulting in an incoherent random walk and rapid convergence to Eq.~\eqref{eq:traj-diff-aniso}.
At $\Delta=1$, heavy quasiparticles retain a parametrically larger bare velocity and can undergo rare but substantial displacements through collisions with other large, mobile species.
At short times, the dynamics of quasiparticles of species $s$ are dominated by collisions with quasiparticles of species $s' < s$, where the dominant $s'$ grows over time as more collisions occur. The associated scattering shift $2s'$ is time-dependent, giving rise to anomalous transport. Eventually, at late times, the dominant scattering events involving species $s$ are with species $s' > s$. The associated scattering shift $2s$ is the maximum one possible for species $s$, so the dynamics in this regime is a random walk with a step size $\sim 2s$.
This provides a scaling interpretation of the KPZ-like term in Eq.~\eqref{eq:traj-diff-iso}; a first-principles derivation of the full fluctuation law remains an open problem.

\begin{figure*}[t!]
  \centering
    \includegraphics[width=1\linewidth]{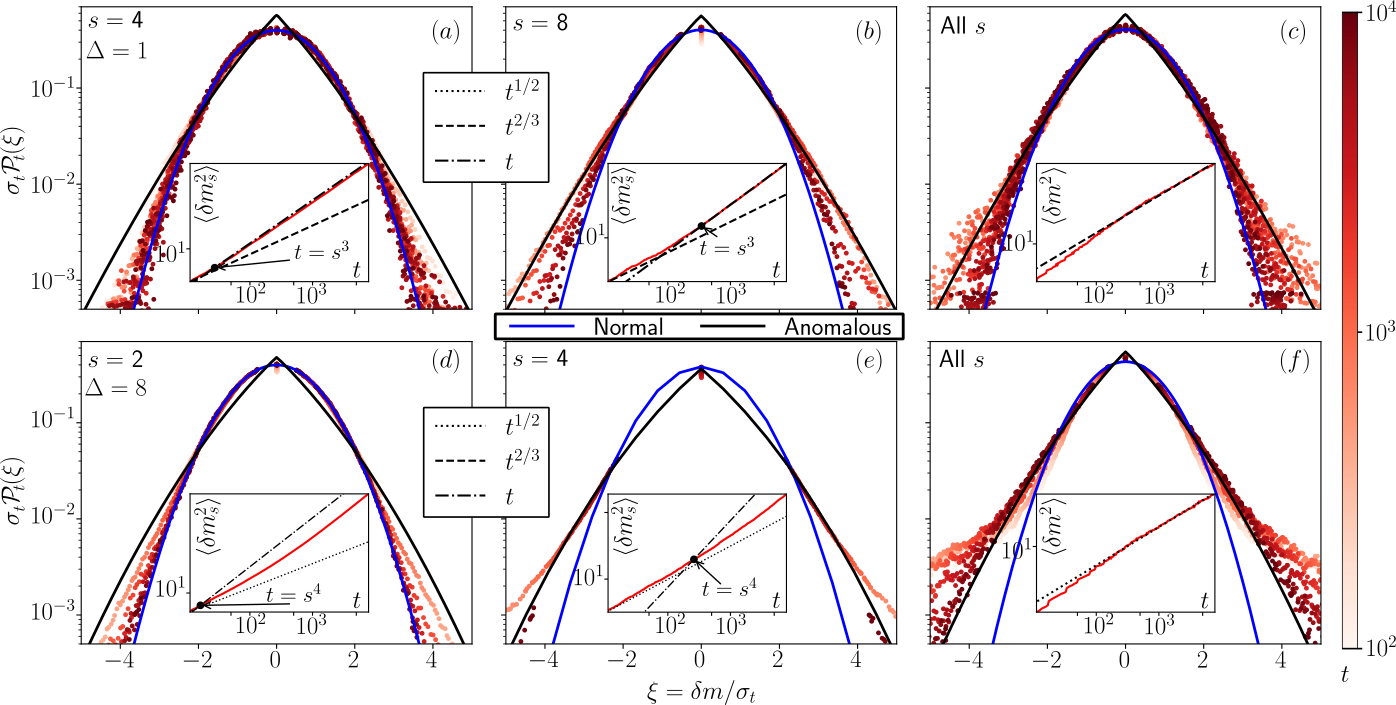}
  \caption{Rescaled distributions of magnetic fluctuations for fixed species $\mathcal{P}_t(\delta m_s)$ and for the total magnetization transfer $\mathcal{P}_t(\delta m)$, Eq.~\eqref{eq:dm-def}, for $\Delta = 1$ (top) and $\Delta = 8$ (bottom). The curves are compared with the anomalous distribution of Eq.~\eqref{eq:anomalous} and with a Gaussian, shown respectively by the solid blue and solid black reference curves. All distributions are normalized to unit variance (the latter indicated by $\sigma_t$). Insets show the corresponding variances as functions of time.}
  \label{fig:magnetization_distribution}
\end{figure*}

\section{Charge transport}
\label{sec:screening}
Fluctuations of particle trajectories are fundamentally different from the fluctuations of the charge they transport. We now turn to charge fluctuations in thermal equilibrium, focusing on the time variation of the total charge contained in the left half-system,
\begin{equation} \label{eq:dm-def}
    \delta m(t)=\sum_{s \geq 1} \delta m_s(t),
    \quad
    \delta m_s(t)=\sum_{x<0} s \bigl(N_s(x,t)-N_s(x,0)\bigr) ,
\end{equation}
with $N_s(x,t)=\sum_i\delta(x-x_{s,i})$. 
The total magnetization fluctuations therefore decompose as
\begin{equation}
    \big\langle \delta m(t)^2\big\rangle
    =
    \sum_{s \geq 1} \big\langle \delta m_s(t)^2\big\rangle
    +
    \sum_{s\neq s'} \big\langle \delta m_s(t)\,\delta m_{s'}(t)\big\rangle.
    \label{eq:dm2-decomp}
\end{equation}
Unlike previous works, our model gives direct access to the species-resolved contributions $\delta m_s$, allowing us to identify the crucial role played by inter-species correlations.

First we should notice that by using Eq.~\eqref{eq:traj-diff-iso}, we find for the time-averaged spin diffusion constant (see End Matter for more details)
\begin{equation}\label{eq:diffusionsdiverg}
\bar{\mathfrak{D}}(t)  \propto  { \int_0^t dt'  \sum_{s \geq 1}   \frac{ \rho_s \ |(v^\sigma_s)^{\rm eff}|^2 s^4 }{\Big|(v^\sigma_s)^{\rm eff} t +   \sqrt{\kappa_{\rm G} s t} H_{\Delta = 1}(t/s^3)   \Big|  } } \sim t^{1/3},
\end{equation}
which gives the KPZ dynamical exponent $z=3/2$, as previously observed numerically and derived from kinetic arguments \cite{GopalakrishnanVasseur2019KT,PhysRevLett.127.057201}. Remarkably, this exponent is controlled only by the crossover scale $t_*(s)\sim s^3$, not by the detailed form of the scaling function $H_{\Delta=1}$.

We now consider the dynamics of each $\delta m_s$. Although all quasiparticles move ballistically and carry positive charge, the fluctuations $\delta m_s(t)$ are not simply ballistic at all accessible times. For each species there is a crossover, see Fig.~\ref{fig:screening_magnetization}, at a parametrically large time $t_*(s)\sim s^{2z}$, with
\begin{equation}
\langle(\delta m_s(t))^2\rangle \sim
\begin{cases}
t^{1/z}, & t\ll t_*(s),\\[2pt]
v_s^{\rm eff}\, t + O(t^{1/z}) & t\gg t_*(s).
\end{cases}
\end{equation}
while the total magnetization remains diffusive/superdiffusive at all times, $\langle(\delta m(t))^2\rangle \sim t^{1/z}$; see Fig.~\ref{fig:magnetization_distribution}. 
Crucially, it is the off-diagonal terms in Eq.~\eqref{eq:dm2-decomp}  which cancel the ballistic contributions of the diagonal ones (see also End Matter for more details): once a given species becomes ballistic, its effect on the \emph{total} magnetization is screened by strong inter-species correlations induced by the scattering shifts. This mechanism controls not only the variance, but the full counting statistics of charge transfer.

\subsection{Anisotropic regime: $\Delta>1$}
\label{subsec:screening-aniso}
For $\Delta>1$, the largest strings have exponentially small bare velocities and are effectively immobile. Their contribution is therefore not screened out even for $t\gg t_*$ (as they lack ballistic transport), but their charge fluctuations are still anomalous: rather than ordinary Brownian motion, they evolve through a stochastic velocity field $v_\infty$ generated by collisions with light particles. At coarse-grained scales the magnetization obeys
\begin{equation}
    \partial_t m + \partial_x\!\left(v_\infty\, m\right)=0,
    \label{eq:meq}
\end{equation}
with $\langle v_\infty\rangle=0$ in equilibrium. This leads to (see End Matter for more details) the anomalous full counting statistics, see \cite{Krajnik2022,PhysRevLett.128.090604,GopalakrishnanMcCullochVasseur2024PNAS,mcculloch2024,krajnik2024_dynCrit,yoshimura_anomalous_2024,2602.24242}, 
\begin{equation}\label{eq:anomalous}
P_t(Q) = \int_{-\infty}^{\infty } dx_\infty \,\mathcal{P}_t(x_\infty)\, \frac{{\rm e}^{- Q^2/(2 \chi |x_\infty| )}}{\sqrt{2 \pi \chi |x_\infty|}},
\end{equation}
where $\chi$ is the magnetic susceptibility and $\mathcal{P}_t(x_\infty)$ is the distribution of the effective displacement of the largest strings, taken Gaussian with variance $2D_\infty t$. This form describes both $\delta m_s$ for sufficiently large $s$ and the total $\delta m$; see Fig.~\ref{fig:magnetization_distribution}. The spin structure factor can likewise be related to the probability distribution of $x_\infty(t)$, up to convolution with the nontrivial equilibrium correlations of the soliton gas, see Fig.~\ref{fig:mxtm00_correlation}.

\subsection{Isotropic point: $\Delta=1$}
\label{subsec:screening-iso}
At $\Delta=1$, there is no longest quasiparticle species. Instead, each string of length $s$ undergoes a crossover at a characteristic time $t_*(s)\sim s^3$, separating an early-time KPZ regime from a late-time ballistic regime with diffusive broadening. For $t\ll t_*(s)$, the distribution of $\delta m_s$ retains the same anomalous form as in Eq.~\eqref{eq:anomalous}, with $x_\infty$ replaced by $x_s$, and with a variance growing as $t^{2/3}$, consistently with Eq.~\eqref{eq:traj-diff-iso}; see Fig.~\ref{fig:magnetization_distribution}(a,b). At later times, $\delta m_s$ crosses over to ballistic transport and its distribution accordingly becomes Gaussian.

A similar crossover is also observed for the \emph{total} magnetization, see Fig.~\ref{fig:magnetization_distribution}(c), although the underlying mechanism is different. Because of screening, at time $t$ only species with $s\gg t^{1/3}$ still contribute, while smaller strings have already entered the ballistic regime and therefore drop out of the sum. At the same time, larger strings begin to contribute to the total charge fluctuations when $v_s t\sim \ell$, where $\ell\sim s^2$ is their typical interparticle distance, which again gives $t\sim s^3$.

As time increases, $\delta m$ therefore becomes a sum over an increasing number of weakly correlated species contributions. The correlations between contributions with $s=\tilde s\, t^{1/3}$ decay algebraically as $t^{-1/3}$ (Fig.~\ref{fig:correlationsdeltams}), in sharp contrast with the anisotropic regime, where the relevant large-$s$ modes remain essentially identical and fully correlated. This produces an effective central-limit behavior for the total magnetization distribution, explaining why it appears close to Gaussian at accessible times even while KPZ scaling is visible in trajectory fluctuations.

\section{Conclusion}
We have introduced a multi-species soliton gas with only two chiral velocities that captures the phase diagram of XXZ spin dynamics at the level of the transport observables studied here. Despite its simplicity, it reproduces diffusion in the anisotropic regime, superdiffusion at the isotropic point, screening of ballistic magnetization transport, and the emergence of nearly Gaussian charge-transfer fluctuations at accessible times. It also reveals an additional manifestation of KPZ physics, namely in the fluctuations of single large-quasiparticle trajectories.

Several questions remain open. Most notably, although our numerics are consistent with KPZ statistics for tagged-particle trajectories, we do not yet have an analytic derivation. It is also striking that KPZ scaling appears both at the level of individual trajectories and in the net magnetization transport, and the precise relation between these two phenomena remains to be understood. We expect that the simplicity of the present model will make it a useful setting for future analytical work, for instance with ballistic macroscopic fluctuation theory \cite{10.21468/SciPostPhys.15.4.136,yoshimura_anomalous_2024,2504.09201,2505.18093,2602.24242}.
\begin{acknowledgments}
\section{Acknowledgments} 
We thank Pierre Le Doussal, Stefano Olla, Alvise Bastianello, for useful comments and Jitendra Kethepalli, {\v{Z}}iga Krajnik and Benjamin Doyon for discussions and collaborations on related topics. 
J.D.N. and A.U. are funded by the ERC Starting Grant 101042293 (HEPIQ) and the ANR-22-CPJ1-0021-01. R.V.~acknowledges support from the Swiss National Science Foundation (grant 10008234). This work was granted access to the HPC resources of IDRIS under the allocation AD010613967R2 and AD010613967R1.
\end{acknowledgments}


\input{xxz_bbl.bbl}

\clearpage
\appendix 

\twocolumngrid
\begin{center}
    \textbf{\large End Matter}
\end{center}

\section{Supplemental numerical data}

\begin{figure}[h!]
  \centering
    \includegraphics[width=1\linewidth]{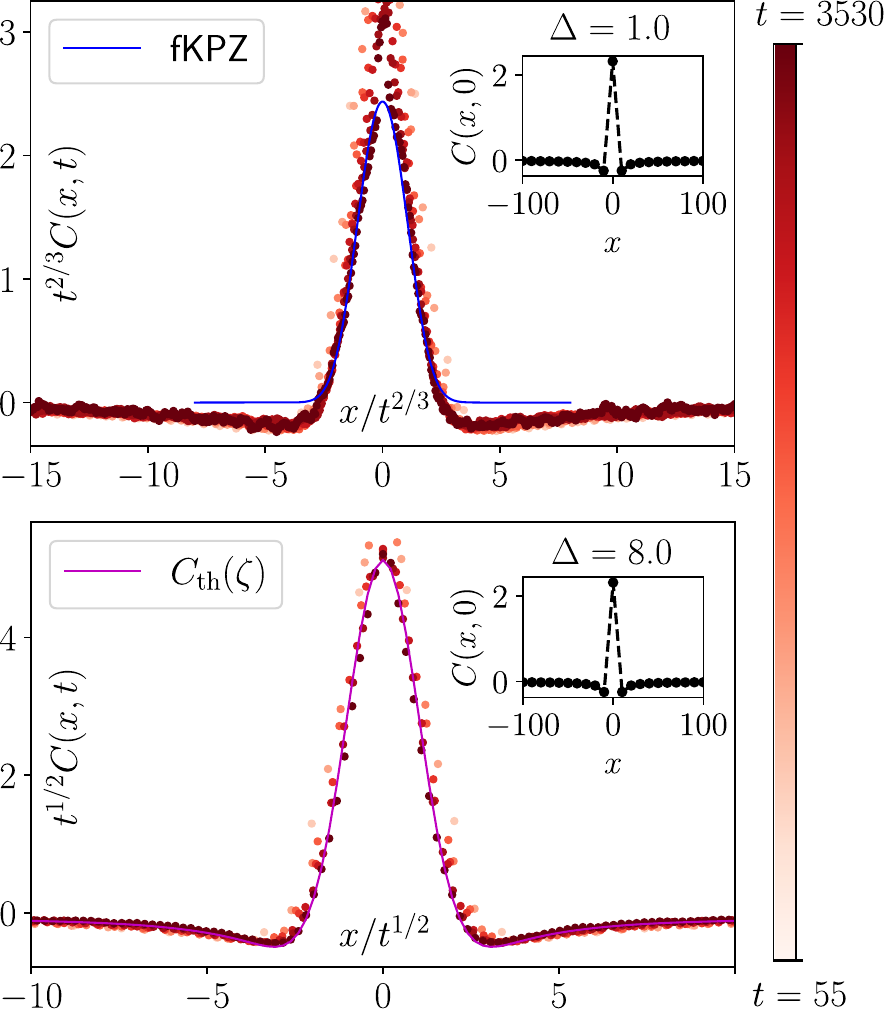}\\
  \caption{  Rescaled correlation function $C(x,t) = \langle m(x,t) m(0,0) \rangle_{\rm c}$ for both $\Delta = 8$ and $\Delta = 1$. In either anisotropy, the scaling is significantly different between the anisotropies. For $\Delta = 8$ the evolution is well described by convolving the initial correlation with the diffusion kernel (continuous line). For $\Delta=1$ the correct convolution is not known, and we show the disagreement with the (naive) distribution $f_{\rm KPZ}$, which is a rescaled Pr\"ahofer-Spohn scaling function~\cite{prahofer_2004}. The insets show the value of the equilibrium correlation at $t=0$ showing that it is not a $\delta(x)$ function.  }
  \label{fig:mxtm00_correlation}
\end{figure}

\begin{figure}[h!]
  \centering
  \includegraphics[width=1\linewidth]{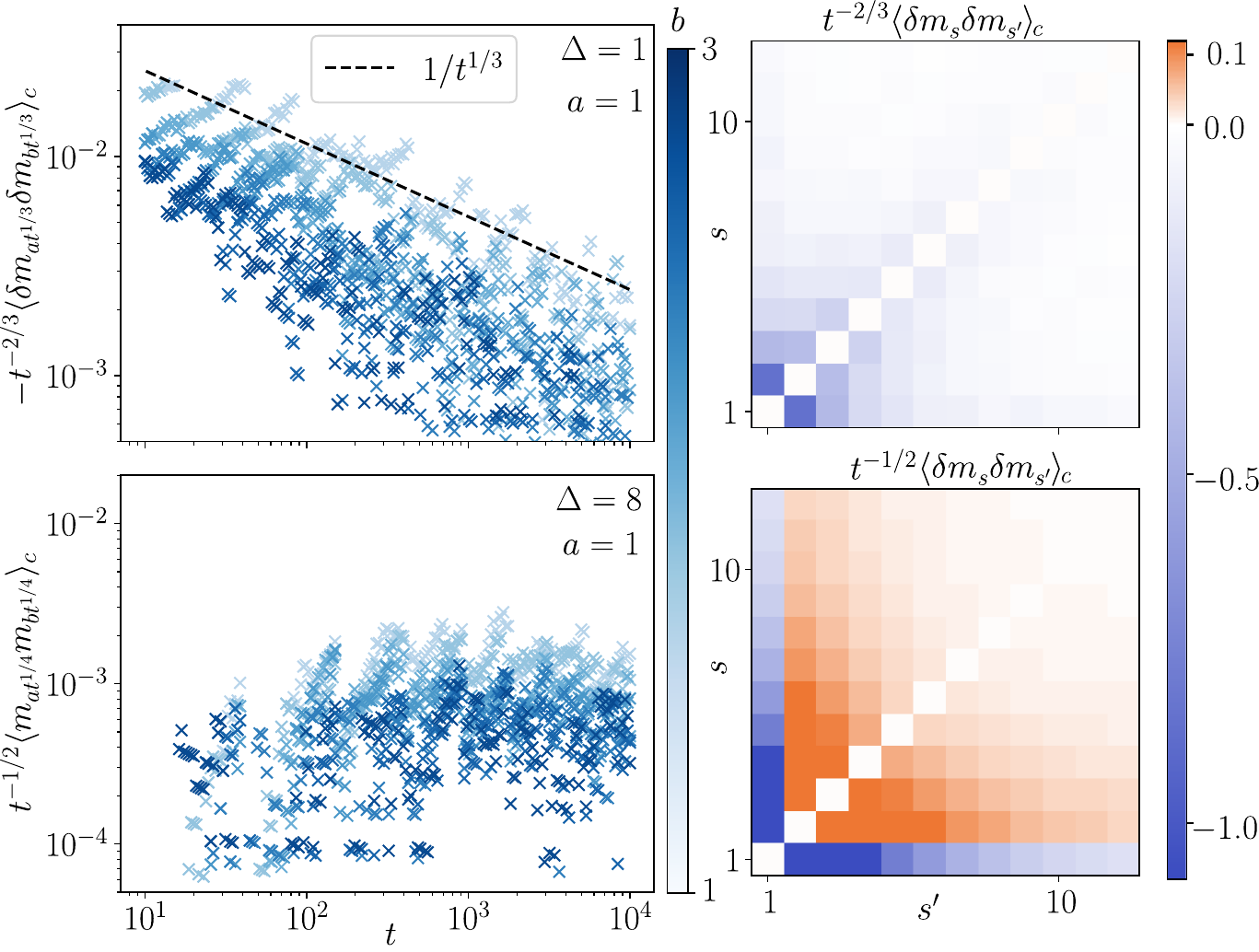}\\
  \caption{ The evolution of the off-diagonal part of the correlation function. In the left plot, the off-diagonal correlation function, divided by the diagonal strength, is plotted  with rescaled spins $s = a t^{1/2 z}$ and $s^\prime = b t^{1/2z}$, where  $a = 1$ and $b$ is changed, with darker blue implying larger $b$. The rescaled correlation function is observed to decay in time as $t^{-1/3}$ when $\Delta = 1$, implying the validity of the CLT, whereas for $\Delta =8$ the rescaled correlation function is constant, suggesting the failure of the CLT. In the second column, the time-scaled off-diagonal part of the correlation functions is plotted at $t \sim 8000$. From these plots the $\Delta = 8$ correlation is found to be dominated by the magnon-string interactions, and it is uniformly non-zero at large particles, while for $\Delta =1$ it vanishes for large particles. }
  \label{fig:correlationsdeltams}
\end{figure}

\begin{figure}[h!]
  \centering
  \includegraphics[width=1\linewidth]{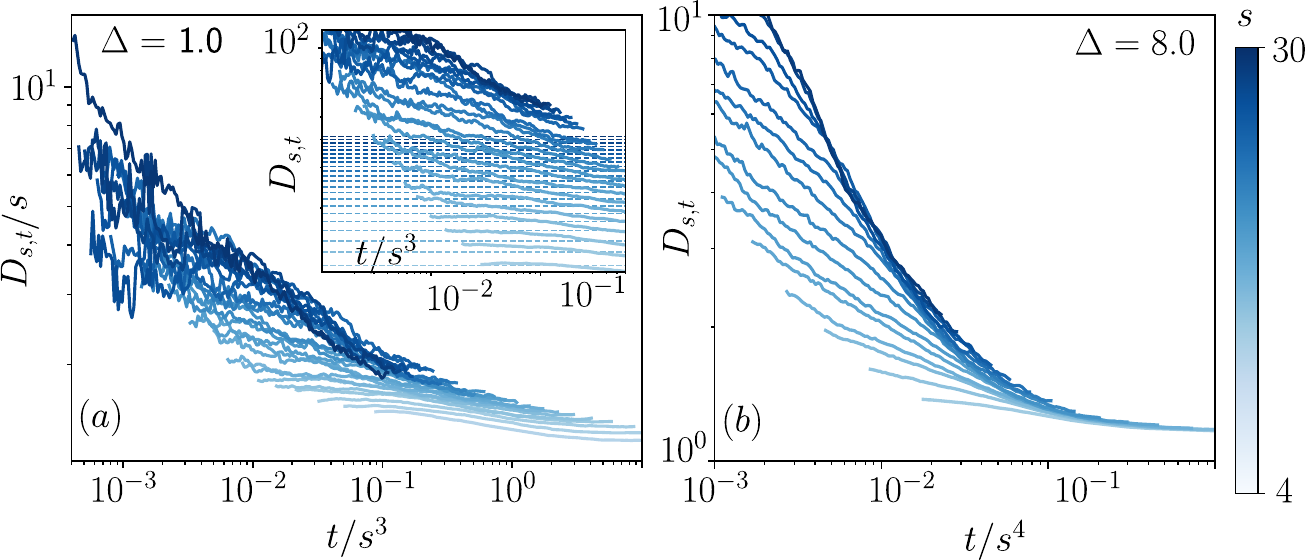}\\
  \caption{ Plot of finite-time diffusion constant $D_{s,t} = \langle (\delta x_{s}(t) - \langle \delta x_{s}\rangle )^2 \rangle/ t$ against rescaled time for each particle species. In plot (a) the diffusion constant is rescaled by the string length with the expectation that $D_{s,t} / s \overset{s \to \infty}{\sim} \mathrm{const}$ and the result demonstrates the collapse of large particles to a single curve; the inset shows the unscaled values and dashed lines indicating the asymptotic diffusion, see Eq.~\eqref{eq:diffusion_exact}. In plot (b) all particles are found to approach the same diffusion constant as expected. }
  \label{fig:dxdx_no_subtraction}
\end{figure}

\begin{figure}[h!]
  \centering
  \includegraphics[width=1\linewidth]{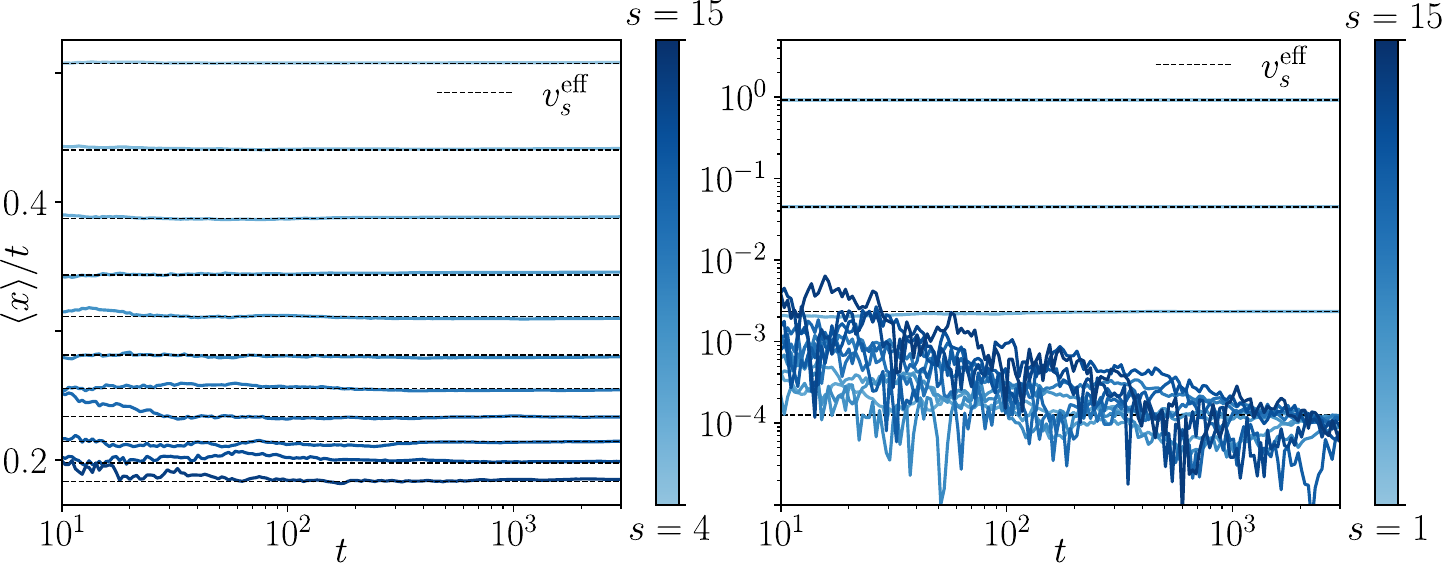}\\
  \caption{ The analytical value of $v^{\rm eff}_s$  (defined in Eq.~\eqref{eq:veff-def}) is plotted as a dotted  line over the numerically determined velocity for each string type, evaluated as $\langle  x_{s,i}(t) \rangle / t$ where $x_{s,i} = x_{s,i}(t) - x_{s,i}(0)$ for $\Delta = 1$ (left) and $\Delta = 8$ (right).}
  \label{fig:velocity_approach}
\end{figure}

\begin{figure}[h!]
  \centering
  \includegraphics[width=1\linewidth]{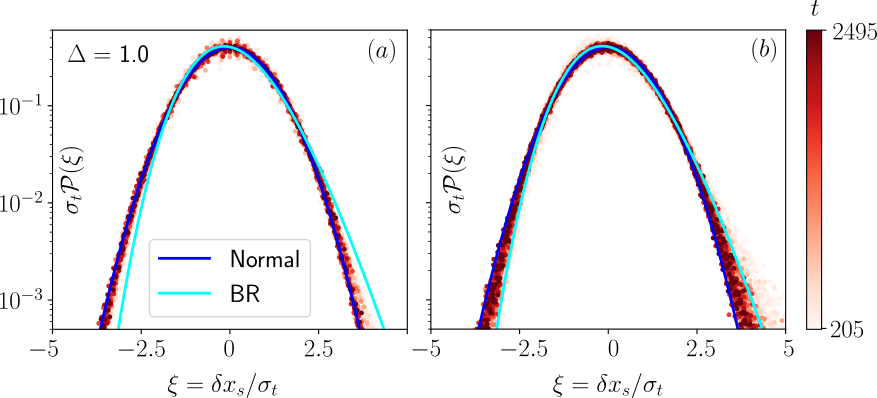}\\
  \caption{ Plot of the distribution of fluctuations $\mathcal{P}(\delta x_s)$ (also averaged with left movers $\xi \to - \xi$) for $\Delta = 1$. The left panel, plot (a), shows the smallest strings, $s = 1 - 5$, whose fluctuations are consistent with a Gaussian distribution. In the right panel, plot (b), intermediate sized strings, $s = 6 - 10$, are plotted and demonstrate a transition from a Baik--Rains distribution to a Gaussian distribution. These distributions are rescaled to have unit variance and zero mean.  }
  \label{fig:smallStringPDF}
\end{figure}
\section{Vanishing ballistic part of magnetization}

In the large anisotropy limit there are two distinct spin types, the ballistic particles $s = 1$, and the immobile particles $ s > 1$. The magnetization fluctuation is noted as being closely associated with the number fluctuation
\begin{eqnarray}
    \delta m_s = s \, (N^L_s - N^R_s) = s \, \delta N_s ,
\end{eqnarray}
where $\rho_s =  \frac{1}{ s (s+1)(s+2)} $. The number fluctuation of an immobile string is proportional to $ - \delta N_1 $, since an excess of right moving $s = 1$ particles induces a leftward motion of the immobile string. This proportionality can be fixed by noting that any immobile string - magnon collision leads to a shift of $\mathfrak{a}_{1,s} = 2$, whereas magnon-magnon collisions lead to shifts $\mathfrak{a}_{1,1}=1$. This shift can be combined with the likelihood of a ballistic - immobile string collision ( $\langle N^{\rm coll}_s \rangle = t v_1 \rho_s$) to imply that the ballistic fluctuations for the immobile particles are given by
\begin{eqnarray}
    \frac{\delta N_1}{\rho_1} = - \frac{2 \delta N_s}{\rho_s}. 
\end{eqnarray}
These ballistic-scale identities are inserted into the total magnetization fluctuation, which after some algebra is rewritten as
\begin{eqnarray}
    \langle \delta m^2 \rangle = \langle \delta m_1^2 \rangle \left( 1 - \sum_{s > 1} s \frac{\rho_s}{2 \rho_1} \right)^2 + O(\sqrt{t})
\end{eqnarray}
The evaluation of the sum in the brackets, $\sum_{s > 1} s \frac{\rho_s}{2 \rho_1}  = \sum_{s > 1} \frac{ 3}{(s+1)(s+2)} = 1$, exactly cancels the bracketed term and leaves only the $O(\sqrt t)$ terms, thereby ensuring that the magnetization is not ballistic. It is worth pointing out is that this cancellation relies on the form of the density and implies that away from half filling ballistic growth of the magnetization variation can occur.

\section{Derivation of the spin diffusion constant}. {
Starting from $ m(x,t)\approx m_0\!\left(x-\int_0^t dt'\,v_\infty(x,t')\right)$, we find that the spin diffusion constant is given by the variance $\langle x_\infty^2 \rangle = 2\mathfrak{D} t +\dots$ of the quantity $x_\infty(t) = \int_0^t dt v_\infty(0,t)$. The average $\langle v_\infty(x,t) \rangle = 0$ vanishes for $\Delta \geq 1$, but there are non-trivial space-time fluctuations due to thermal fluctuations in the initial state. Within linear response, note that 
\begin{equation}
\delta v_\infty = \sum_{s,\sigma}  \frac{\delta v_\infty }{\delta n_{s,\sigma}} \delta n_{s,\sigma} + \dots  
\end{equation}
where the functional derivative is evaluated in the background equilibrium state, whereas the fluctuations $\delta n_{s,\sigma} (x,t)$ are simply advected from the initial state along the particle trajectories. We then define a time-dependent diffusion constant diffusion constant from $\bar{\mathfrak{D}}(t) = \frac{1}{2} \frac{d}{dt} \langle x_\infty^2(t)\rangle$, so that
\begin{align} 
\bar{\mathfrak{D}}(t) & = \frac{1}{2} \frac{d}{dt} \lim_{s_0 \to \infty} \int_0^t dt'  \int_0^{t} dt''  \sum_{s,s'} \sum_{\sigma,\sigma'} \frac{\delta v_{s_0} }{\delta n_{s,\sigma} }   \frac{\delta v_{s_0} }{\delta n_{s',\sigma'} } \nonumber  \\& \langle  \delta n_{s,\sigma} (x_s(t))\delta n_{s', \sigma'} ( x_{s'}(t) )  \rangle,
\end{align}
with the convective particle trajectories following eq.~\eqref{eq:traj-diff-iso}. Using the thermal fluctuations in the initial state
\begin{equation} 
 \langle \delta n_{s, \sigma } (x)\delta n_{s' ,\sigma'} (x')  \rangle = \frac{\rho_{s} }{(\rho_{s }^{\rm tot} )^2 } \delta_{s,s'} \delta(x-x') \delta_{\sigma,\sigma'},
\end{equation}
and using that }
\begin{equation}
     \lim_{s_0 \to \infty}  \left( \frac{\delta v_{s_0} }{\delta n_{s,\sigma} }\right)^2 \propto s^4 ((v^\sigma_s)^{\rm eff}  \rho_{s}^{\rm tot} )^2,
\end{equation}
we find~\eqref{eq:diffusionsdiverg} in the main text. 

\section{Single-species limit and hard-rod contraction}
A better intuition for the role of the trajectories can be gained by restricting ourselves to a single particle species. In this case, the scattering kernel 
\begin{eqnarray}
    \mathfrak{a}_{s,s'} = 2 \min(s,s') - \delta_{s,s'} \overset{s = 1}{\to} 1 = \mathfrak{a}.
\end{eqnarray}
With this identification the auxiliary coordinates, characterized by free point particles, reproduce the hard-rod model by fixing $s = 1$, yielding hard rods with length $ \mathfrak{a} =1$. Since there is only a single particle species it follows that 
\begin{eqnarray}
 X^{(t)}_{s,i} &=& x_{s,i}(t) + \frac 1 2 \sum_{s',j} \mathfrak{a}_{s,s'} \, \mathrm{sign}\left( x_{s,i}(t) - x_{s',j}(t)\right) \nonumber \\
    X^{(t)}_{i}  &=& x_{i}(t) + \frac {\mathfrak{a}} 2 \sum_{j}  \, \mathrm{sign}\left( x_{i}(t) - x_{j}(t)\right) \, .
\end{eqnarray}
This last line is exactly the contraction map for hard rods~\cite{Spohn1991}. The multi-species model studied in the main text does not reduce to a geometric hard-rod gas, because the scattering shift in Eq.~\eqref{eq:scatteringT} is nonlinear in the two species labels.

\section{Magnetic fluctuations in the anisotropic regime: $\Delta>1$}
Here the largest particles with $s\to \infty$ have exponentially small bare velocities and remain effectively immobile. As a consequence their contribution to $\delta m(t)$ is not screened out.
Instead, their charge evolves according to a stochastic velocity $v_\infty$ that fluctuates due to scatterings with light particles.
At the coarse-grained level, therefore, the magnetization density obeys an effective continuity equation
\begin{equation}
    \partial_t m + \partial_x\!\left(v_\infty\, m\right)=0.
    \label{eq:meq_EndMatter}
\end{equation}
In equilibrium $\langle v_\infty\rangle=0$, but $v_\infty$ exhibits non-trivial space--time fluctuations inherited from the initial (thermal) configurations.
Solving Eq.~\eqref{eq:meq_EndMatter} by characteristics for an initial profile $m(x,0)=m_0(x)$ gives $m(x,t) = \int dx_0\, m_0(x_0)\,\delta(x-x_\infty(x_0,t))$, where the characteristic position of the large particle $x_\infty(x_0,t)$ satisfies $\partial_t x_\infty = v_\infty(x_\infty,t)$ with $x_\infty(x_0,t=0)=x_0$.
Since $v_\infty$ is random with short-ranged temporal correlations at fixed $x$, the characteristic is expected to be diffusive, given by Eq.~\eqref{eq:traj-diff-aniso} with $s \to \infty$.
At long times, we may therefore evaluate the advected fields on the mean (Euler) trajectory (here corresponding to zero drift velocity), effectively replacing $v_\infty(x_\infty,\tau)$ by $v_\infty(x,\tau)$.
This leads to the approximation $ m(x,t)\approx m_0\!\left(x-\int_0^t dt'\,v_\infty(x,t')\right)$. 
This equation can be used to evaluate various quantities, including full counting statistics, which is given in terms of an anomalous, non-normal distribution~\cite{GopalakrishnanHuseKhemaniVasseur2018PRB,Krajnik2022,PhysRevLett.128.090604,GopalakrishnanMcCullochVasseur2024PNAS,yoshimura_anomalous_2024}, expressed as a product of two random variables,
\begin{equation}\label{eq:anomalous_EndMatter}
P_t(Q) = \int_{-\infty}^{\infty } dx_\infty \,\mathcal{P}_t(x_\infty)\, \frac{{\rm e}^{- Q^2/(2 \chi |x_\infty| )}}{\sqrt{2 \pi \chi |x_\infty|}},
\end{equation}
one coming from the fluctuations of the initial state in terms of the magnetic susceptibility $\chi$ (e.g.\ $P(m_0)\sim e^{-(m_0)^2/(2\chi)}$), and $\mathcal{P}_t(x_\infty)$ taken to be normal with zero mean and variance $2 D_\infty t$, giving the fluctuations of $v_\infty$.
This distribution describes both the fluctuations of $\delta m_s$ with $s \gg 1$ and of the total charge $\delta m$; see Fig.~\ref{fig:magnetization_distribution} (e,f) as all the $\delta m_s$ for sufficiently large $s$, fluctuate in the same exact way.

Moreover, the \textit{charge (spin) structure factor} can be related to the probability distribution of $x_\infty(t)=\int_0^t dt'\,v_\infty(x,t')$: $C(x,t)=\langle m(x,t)\, m(0,0) \rangle  \propto \langle C_0 (x-x_\infty(t)) \rangle$, where $C_0(x)$ is the profile of the initial correlations.
When the latter is a $\delta(x)$ function, the structure factor is precisely the distribution of the fluctuations of the largest-particle trajectory, which is purely Gaussian.
In this soliton gas model, however, equilibrium correlations have a non-trivial structure: convolving these with a Gaussian gives the correct result; see Fig.~\ref{fig:mxtm00_correlation}.

\section{Details on the numerical procedure}
For our algorithm our input parameters were selected as $\bar \rho_s = 1 / (s (s+1) (s+2) )$, $s_{\rm max} = 2000$, and the anisotropy values $\Delta=1, 8$, wherever notated. Conceptually, our method allows us to easily reach any time however, we must fix the particles to be contained within a box $L = 16000$, in units of string length, thereby ensuring constant density, although erroneous correlations develop and saturate at large enough times. The code and some sample data can be accessed online~\cite{urilyon_code} the key steps are described in detail below.\\

The broad sketch of the numerics used to generate the data in this paper follows the algorithm 
\begin{itemize}
    \item Prepare an initial state of auxiliary particles.
    \item Invert the contraction map, Eq.~\eqref{eq:traje}, and truncate the initial state of auxiliary particles in interacting coordinates. Save the truncated auxiliary particle coordinates, which now constitute the initial fluctuating state.
    \item Evolve this initial state forward in time in auxiliary coordinates and invert the contraction map to determine $x_{s,i}(t)$, saving these values as trajectories.
\end{itemize}
Each step must be done carefully and is described below.

\subsection{Inverting the contraction map}
Key to the wave-packet gas method is to take the auxiliary coordinates $X_{s,i}$ and determine the corresponding interacting coordinates $x_{s,i}$ by inverting
\begin{eqnarray}
\label{eq:jacobian}
    X^{(t)}_{s,i} = x_{s,i}(t) + \frac 1 2 \sum_{s',j} \mathfrak{a}_{s,s'} \mathrm{sign} (x_{s,i}(t) - x_{s',j}(t) )
\end{eqnarray}
directly. This is not particularly feasible, except in the case of only one particle species. Instead, consider a surface fixed by the auxiliary coordinates with dependence on a continuous set of coordinates $\tilde x$ and whose form is defined so that its minimum yields the contraction map
\begin{equation}
\label{eq:convexSurface}
    S(\tilde x) := \sum_{s,i}\frac{(X_{s,i} - \tilde x_{s,i})^2}{2} +  \sum_{(s,i),(s',j)} \mathfrak{a}_{s,s'} \frac{|\tilde x_{s,i} - \tilde x_{s',j}|}{4} \, .
\end{equation}
By finding the solution $(x_1  \dots x_N)(t) = \mathrm{Argmin}(S_{X,t}(\tilde x))$ for this convex surface, the contraction map is inverted. For the positive scattering lengths used here, the quadratic term makes the minimizer unique up to particle coincidences, which are resolved by the regularization below. The numerical output is therefore an approximation to the exact minimizer of Eq.~\eqref{eq:convexSurface}, with the error controlled by the minimization tolerance and the final regularization parameter. It should also be noted that this minimization scheme is only applicable to scattering kernels $\mathfrak{a}_{s,s'}$ defining surfaces whose second derivatives satisfy $S_{X,t}''(x) > 0$.

To aid in numerical minimization, the absolute value and sign functions are regularized with an arbitrary factor $\alpha$ as
\begin{eqnarray}
    | x | &\overset{ \alpha \gg 1}{\sim}& \frac{ \log \left[(1 + e^{\alpha x})(1 + e^{-\alpha x}) \right]}{\alpha}, \\
    \mathrm{sign}(x) &\overset{ \alpha \gg 1}{\sim}& \tanh( \alpha x/ 2)\, .
\end{eqnarray}
For the data used in the plots the minimization was done via an L-BFGS-B algorithm. For the data used in the plots the minimization was done via L-BFGS-B algorithm. The regularization parameter $\alpha = 10^{\beta}$ was increased in equally scaled logarithmic steps (from $\beta = -2, 2$). 
\\

\subsection{Auxiliary particle initialisation:} A target coarse grained interacting density, $\bar \rho_s$, is identified along with a corresponding coarse grained auxiliary density $\bar \rho^F_s = \bar \rho_s / (1^{\rm dr}_s)$, where $1^{\rm dr}_s = 1 - \sum_{s'} \mathfrak{a}_{s,s'} \bar \rho_{s'}$ is a dependent factor that relates the auxiliary to interacting coordinates and indicates the allowed length through which the interacting particles propagate with their bare velocities. Each species is distributed according to a Poisson point process, where the distance $r_{s,i}$ between the nearest particles of the same type is sampled from a normalized exponential distribution $P(r_s) = \rho^F_s e^{-\bar \rho^F_s r}$ along an axis of length $L$. Repeat this for each particle type $s$. Once placed, a particle is assigned a rapidity (or momentum), in our model simply $\pm 1$; more general assignments can also be made. 

\subsection{Truncate the initial state:} The contraction map is inverted for the combined (multi-species  $s = 1, 2 \dots$) coordinates to determine the interacting coordinates $x_{s,i}$. A truncation is made on the newly obtained interacting coordinates so that they all fall within a region $x_{s,i} \in (-L/2,L/2)$, and the corresponding truncated set of auxiliary coordinates $X_{s,i}$ is saved.


\subsection{Time evolution}
The system is now straightforwardly evolved. In the auxiliary coordinates the particles have a velocity given by $v_s$ with the time evolution
\begin{eqnarray}
    X_{s,i}^{(t)} = v_s t + X_{s,i}^{(0)}
\end{eqnarray}
The time value is arbitrary and the corresponding $x_{s,i}(t)$ is determined by inverting the contraction map.\\

To reach longer times reflective boundary conditions can be introduced. This must be implemented in the auxiliary coordinates to correctly account for reflection, so the reflective boundary will have different auxiliary-coordinate positions for different particle species,
\begin{eqnarray}
    \pm \frac {L^{\rm ref}_s} 2 = \pm \frac L 2 \pm \frac 1 2 \sum_{s',j} \mathfrak{a}_{s,s'} .
\end{eqnarray}
The time evolved auxiliary coordinates are reflected back into the box as
\begin{eqnarray}
    X^{(r)}_{s,i}(t) = \frac{L^{\rm ref}_s}{2} - | \mathrm{ mod}_{2 L^{\rm ref}_s} [X(t)]- L_s^{\rm ref} | .
\end{eqnarray}
This is combined with a reflection number that counts for the number of reflections and combined with the above determines the total displacement of a particle.

\section{Exact solutions of thermodynamic functions}
The simple form of the particle-particle scattering kernel in our model allows for the explicit computation of many relevant transport quantities. Key here is the notion of dressing, in which the single particle variable is modified by surrounding particles into an effective form. To determine the effective velocity a particle's spatial displacement must be rescaled by the portion of space that supports particle presence, a so-called allowed length; this factor is denoted here as $1^{\rm dr}$. This allowed length factor is computed by summing all particles to remove the forbidden regions as
\begin{eqnarray}
    1^{\rm dr}_s = 1 - \sum_s \mathfrak{a}_{s,s' } \bar \rho_{s'}  
\end{eqnarray}
with $\bar \rho = 1 / ( s (s+1) (s+2) )$ being the coarse grained particle density also introduced in the text. After inserting the scattering kernel $\mathfrak{a}_{s,s'} = 2 {\rm min}(s,s') - \delta_{s,s'}$ the sum evaluates to
\begin{eqnarray}
1^{\rm dr}_s =   1 - \left( \frac{s^2 + s - 1}{s (2+s)} \right) = \frac{s+1}{s (s+2)}.
\end{eqnarray}
In our model, the effective velocity is found by rescaling the particle displacement by the allowed length 
\begin{eqnarray}
    (v_s^\sigma)^{\rm eff} = \lim_{t \to \infty} \frac{1}{1_s^{\rm dr} t} \langle v^\sigma_s t\rangle = \frac{s(s+1)}{s+2}v_s^\sigma
\end{eqnarray}
yielding the form used in the text.

\subsection{General approach for dressing calculations}
Generically, other variables can be understood as being 'dressed' by the identity
\begin{eqnarray}
    \sum_{s'} ( \delta_{s,s'} +   \mathfrak{a}_{s,s'} n_{s'} ) f^{\rm dr}_{s'} = f_s \, ,
\end{eqnarray}
where the new variable $n_s = \bar \rho_s / 1_s^{\rm dr} = 1 /( s+1)^2$ is obtained by comparison with our calculation for the allowed length. The goal now is to rewrite this expression in a recursive form. The first step is to insert the scattering kernel and grouping terms
\begin{eqnarray}
f_s &=& (1 - n_s) f^{\rm dr}_s + 2\sum_{s'=1}^s s' n_{s'} f^{\rm dr}_{s'} + 2 s \sum_{s' = s+1} n_{s'} f^{\rm dr}_{s'} \nonumber\\
f_s^{\rm dr} &=& \frac{ f_s - 2 F_s}{1 - n_s} \, .
\end{eqnarray}
By straightforward substitution a recursive formula for $F_s$ is obtained
\begin{eqnarray}
    F_s &=& \sum_{s = 1}^{s-1} ( s' - s) n_{s'} f_{s'}^{\rm dr} + s \sum_{s'=1} n_{s'} f^{\rm dr}_{s'} \nonumber\\
                &=& \sum_{s' = 1}^{s-1} ( s' - s) n_{s'} \frac{(f_{s'} - 2 
                F_{s'})}{ 1- n_{s'}} + s \sum_{s' = 1} \rho_{s'} f_{s'}  
\end{eqnarray}
Now, given a function $f_s$, this recursive formula is used to compute the function $F_s$, from which an ansatz is constructed and checked against a difference equation for $F_s$ (following from the above equation). Thus determining the function $F_s$ and fixing the form of the dressed function. \\

More concretely, the diffusive contribution from each particle relies on the dressed form of the scattering kernel. This must be treated with a simple modification $F_s \to F_{s,s'}$, with the note that the bare function is $f_s \to \mathfrak{a}_{s,s'}$. With an additional note that the kernel satisfies the transpose identity $\mathfrak{a}^{\rm dr}_{s,s'} = \mathfrak{a}^{\rm dr}_{s',s}$ and with the above method the explicit form of the dressed scattering kernel is found to be 
\begin{eqnarray}
    \mathfrak{a}^{\rm dr}_{s,s'} = \frac{ (\max (s,s')+1) (\min (s,s')+1)^2}{3\max(s,s') ( \max\left(\frac s 2, \frac{s'}{2} \right) + 1)} - \delta_{s,s'} \frac{ (1+ s)^2}{s(2+s)} .\nonumber\\
\end{eqnarray}
Note that this expression for the dressed scattering kernel allows us to directly calculate any dressed variable.

\subsection{Diffusion formula}
\label{app:analytical_diffusion}
With the analytical form of the dressed kernel and the effective velocity, the diffusion constant can also be determined. We separate the positive and negative components of the effective velocity, writing $v_{s}^{\rm eff} = 1 / (s^2 1^{\rm dr}_s)$, and note the convenient identity $v_s^{\rm eff} > v_{s'}^{\rm eff}$ when $ s < s'$. With this additional simplification, the diffusion constant due to a single particle species is
\begin{widetext}
    \begin{eqnarray}
    D_s &=& \frac{ 2}{(1^{\rm dr}_s)^2} \left(  v^{\rm eff}_s \sum_{s' = {s+1}}^\infty  \bar\rho_{s'} (\mathfrak{a}^{\rm dr}_{s,s'})^2 + \sum_{s'=1}^{s-1} v_{s'}^{\rm eff} \bar\rho_{s'} (\mathfrak{a}^{\rm dr}_{s,s'})^2 + v_s^{\rm eff} \bar\rho_s (\mathfrak{a}^{\rm dr}_{s,s})^2 \right) \nonumber \\
    \label{eq:diffusion_exact}
    &=&   \frac 8 9 \sum_{s' =1}^{\infty} \min\left[1, \frac{(s+2)^3s(s+1)}{(s'+2)^3s'(s'+1)} \right] \left(\frac{  s' +1}{ s'} \right)^2 -\frac{2(1 +  4 s) }{3 s^2} \overset{ s \to \infty}{\sim} \frac{ 10 s}{9}
\end{eqnarray}
An explicit summation is possible, although the closed form has an unwieldy form 
\begin{eqnarray}
D_s  
 =\left(P(s)- \frac{s (s+1) (s+2)^3}{9} \psi ^{(2)}(s+1)+\frac{8}{9} \left(2 H_{s-1}+H_{s-1}^{(2)}\right) \right) \nonumber \\
P(s) := \frac{2}{9 s^2 (s+1)^2}-  \frac{s (s (s (s (2 s (s+8)+29)+25)+28)+28)+8}{18 s (s+1)^2}
\end{eqnarray}
with polygamma functions, $\Psi^{(n)}(z) = \partial_z^{n+1} \ln \Gamma(z)$, and generalized harmonic numbers $H_{s-1}^{(n)} = \sum_{k=1}^{s-1} 1/ k^n$
\end{widetext}

\end{document}

%% file: xxz_bbl.bbl
%